\documentclass[aps, prd, onecolumn, showpacs, superscriptaddress, groupedaddress]{revtex4} 
\usepackage{graphicx}	
\usepackage{amssymb}
\usepackage{dcolumn}
\usepackage[mathscr]{euscript}
\usepackage{amsfonts}
\usepackage{amsmath}
\usepackage{color}
\usepackage{subfigure}
\usepackage[a4paper, total={6.2in, 10in}]{geometry}
\usepackage{epsfig}
\usepackage{commath}
\usepackage{subfigure, rotating, bm, array}
\usepackage[pagebackref=false, colorlinks=true]{hyperref}

\hypersetup{
linkcolor=blue,     
citecolor=blue,     
urlcolor=blue}      

\begin{document}

\title{Tidal disruption of a neutron star near naked singularity}

\author{Ashok B. Joshi}
\email{gen.rel.joshi@gmail.com}
\affiliation{PDPIAS,
Charotar University of Science and Technology, Anand- 388421 (Guj), India.}
\affiliation{International Centre for Space and Cosmology, School of Arts and Sciences, Ahmedabad University, Ahmedabad-380009 (Guj), India.}
\author{Pankaj S. Joshi}
\email{psjcosmos@gmail.com}
\affiliation{International Centre for Space and Cosmology, School of Arts and Sciences, Ahmedabad University, Ahmedabad-380009 (Guj), India.}
\author{Sudip Bhattacharyya}
\email{sudip@tifr.res.in}
\affiliation{Department of Astronomy and Astrophysics, Tata Institute of Fundamental Research, 1 Homi Bhabha Road, Colaba,
Mumbai 400005, India.}
\affiliation{MIT Kavli Institute for Astrophysics and Space Research, 
Massachusetts Institute of Technology, Cambridge, MA, 02139, USA}

\date{\today}

\begin{abstract}
We investigate the tidal disruption of a neutron star (NS) near a black hole (BH) and a naked singularity (NaS). For a BH with a mass greater than about $10 M_{\odot}$, the tidal disruption of an NS should occur within the event horizon, and hence neither can the stellar material escape nor a distant observer observe the disruption. 
Since a NaS does not have an event horizon, a significant portion of the NS's material can escape, and the tidal disruption can be observed by a distant observer.
One could identify such an event from the observed emission from the disrupted NS's material and the decay of the light curve of the disruption event. 
The escape of a significant fraction of the NS's material may also have implications for the heavy elements in the universe.
Moreover, the observation of such an event can be useful to confirm a NaS, to probe its spacetime and the motion of matter in such a spacetime, and also to constrain NS parameters and equation of state models. 
As a first step in this direction, we calculate the tidal disruption radius and other parameters for a specific type (Joshi-Malafarina-Narayan type 1) of NaS and compare our results with observations.  

\bigskip

$\boldsymbol{key words}$: Neutron Star, Tidal disruption events, Naked Singularity, Black hole.
\end{abstract}
\maketitle

\section{Introduction}
The Cosmic Censorship Conjecture (CCC) is one of the most extensively debated issues in gravitation physics today, because of its deep theoretical significance in black hole (BH) physics as well as its astrophysical implications \cite{Hawking,wald,Joshi:2008zz,Joshi:2002dt,Goswami:2004gy,Joshi:2024gog}. The weak cosmic censorship states that any spacetime singularities in the universe can never be visible to faraway observers in spacetime or from future null infinity. According to the strong version, singularities cannot likewise be even locally naked. A family of null geodesics can leave the singularity during gravitational collapse, but they are unable to cross the matter cloud's boundary and return to the singularity. This is referred to as a singularity's local visibility.  It turns out that for non-homogeneous marginally bound dust collapse, there are multiple possibilities for both local and global appearance of a singularity \cite{Joshi:1993zg,Eardley:1978tr}. Several theoretical and observational studies have been carried out to distinguish between BHs and naked singularities (NaSs) \cite{Joshi:2013dva, Joshi2020, Joshi:2023ugm, Chakraborty:2024jma, Tahelyani:2022uxw, Patel:2022vlu, Patel:2023efv, Bambhaniya:2019pbr}. In this study, we consider the tidal disruption radius for NS in BH and NaS spacetimes. Here, we use Schwarzschild and Joshi-Malafarina-Narayan (JMN1) spacetimes \cite{psJoshi1} for BH and NaS, respectively. The causal structure for these geometries is examined in terms of the tidal nature of the BH and the NaS, explicitly deriving the tidal disruption radii in these spacetime models. \\

The phenomenon of stellar tidal disruption by massive black holes was first theoretically proposed by Hills (1975)  \cite{Hills}, with observational implications later explored by Lacy et al.\ (1982) \cite{Lacy}. Rees (1988) summarized these developments and predicted a fallback rate scaling as $\sim t^{-5/2}$ \cite{Rees:1988}. This behaviour was subsequently refined by Phinney (1989) \cite{Phinney}, who showed that the post-disruption accretion rate follows the now-standard $\sim t^{-5/3}$ scaling rate. Early analytic models of stellar disruption and the formation of bound and unbound debris, in a relativistic framework by Evans and Kochanek \cite{Evans:1989qe}. Observational progress in identifying and characterizing tidal disruption events (TDEs) has accelerated significantly in recent years, as summarized in the comprehensive review by Gezari (2021) \cite{Gezari}. The relativistic dependence of the debris energy distribution and the resulting mass fallback rate has been analyzed in detail in \cite{Jankovic:2023nyy}. In partial disruption scenarios, the mass fallback rate can follow alternative power-law scalings, as demonstrated in relativistic analyses of partial stellar disruption \cite{Coughlin:2019pqk}. Large samples of optically, UV-, and X-ray-selected TDEs have been compiled in recent catalogues, including those by Hammerstein et al.\ (2023) \cite{Hammerstein}, Yao et al.\ (2023) \cite{Yao}, and van Velzen et al.\ (2021) \cite{vanVelzen}. These events have been discovered primarily through wide-field transient surveys such as the Zwicky Transient Facility (ZTF), the All-Sky Automated Survey for Supernovae (ASAS-SN), the Panoramic Survey Telescope and Rapid Response
System (Pan-STARRS), and the Sloan Digital Sky Survey (SDSS), among others. Several TDE candidates in the X-ray and UV bands were later confirmed by observational campaigns \cite{Pahari:2024rev,Cocchi:2023mgm}; well-researched examples, like ASASSN-14li, provide comprehensive multi-wavelength data \cite{Holoien:2016}. The sample of known TDEs has been greatly increased in recent years by large optical transient surveys (e.g., ZTF, ASAS-SN), which have shown a greater diversity in luminosity, emission features, and timescales \cite{Komossa:2015,Dai:2018}. Finding faint and rapidly evolving TDEs, like ASASSN-23bd, further raises the possibility of new subclasses and demonstrates how TDEs can be used to study accretion physics and black hole demographics \cite{Hoogendam:2024tnk, Bhattacharya:2025qps, Wevers:2023ksb, Zhou:2025lzg, Pasham:2022oee, Bhattacharyya:2017tos, Ozel:2016oaf}.\\

While black holes represent the most widely accepted outcome of gravitational collapse, general relativity also admits solutions in which spacetime singularities are not hidden behind event horizons. Such naked singularities arise in a variety of collapse scenarios and continue to play a central role in discussions of the cosmic censorship conjecture. Over the years, considerable effort has been devoted to understanding whether these objects can be distinguished observationally from black holes, through phenomena such as gravitational lensing, accretion dynamics, and strong-field signatures \cite{Joshi:2008,Virbhadra:2002,Bambi:2013}. These studies motivate the search for invariant, physically measurable quantities capable of probing the extreme curvature regions associated with horizonless compact objects. According to \cite{Lattimer:1976kbf}, BHs with masses between 8$-$17 solar masses have enough mass to trigger the typical neutron star (NS) disruption that an asymptotic observer could see; the exact range depends on the NS equation of state. A binary NS, a BH-NS, and a NaS-NS binary system coalesce, and this phenomenon is important because it results in high-energy events, such as gamma-ray bursts (GRBs). We now know that short-duration (sGRB) events, a subtype of GRB signals, originate from the cataclysmic merger of binary NSs. The gravitational-wave signal GW170817, which was generated when two NSs spiraled together and collided, was observed by scientists for the first time on August 17, 2017 \cite{LIGOScientific:2017vwq}. This event was unique among BH merger findings, as it was also observed in light, initially as a GRB and later throughout the entire electromagnetic spectrum. For the first time, scientists were able to observe the same cosmic event (kilonova) using both conventional telescopes and gravitational waves. This discovery demonstrated how heavy elements like gold and platinum are created and showed that short gamma-ray bursts are powered by merging NSs \cite{Metzger:2019zeh, Abbott:2017multimessenger, Abbott:2017kilonova, Abbott:2019properties, Abbott:2020gwtc2}.\\

The fate of NS near a compact object is critically dependent on whether the companion is a NaS or a BH. These encounters are classified as tidal disruption events (TDEs), a larger class of events in which the incoming star is torn apart by the powerful gravitational field of a compact object, releasing massive amounts of energy in the form of radiation and outflows. If a BH's mass surpasses a comparatively low threshold, the NS should pass through the event horizon mostly intact, with minimal likelihood of noticeable disturbance. On the other hand, for both stellar-mass and supermassive cases, the tidal breakup of the NS may be visible to distant observers around a NaS, which does not have an event horizon. Interestingly, if a large fraction of the NS material escapes, bright kilonova-like emissions driven by r-process nucleosynthesis could be seen for supermassive NaSs. Depending on how much of the disrupted matter escapes, such events could even add more heavy elements to the universe than normal NS–NS mergers. According to this viewpoint, research on NS disruptions by BHs and NS–NaS systems has implications for element formation, astrophysical transients, and the energy efficiency of accretion processes, in addition to probing strong gravity.\\

Relativistic tidal disruption in Schwarzschild spacetime has been extensively studied since the pioneering analyses of Carter and Luminet \cite{Carter:1982,Carter:1983,Luminet:1986}, which established the role of strong-field tidal forces in stellar deformation and debris energetics. Subsequent work has refined these results within fully relativistic frameworks. A key requirement for any robust probe of strong-field gravity is invariance under coordinate transformations. Curvature invariants and covariantly defined tidal indicators satisfy this criterion, encoding spacetime geometry directly through the Riemann tensor and its contractions, and have proven effective in distinguishing black hole geometries even in extreme relativistic regimes \cite{Cherubini:2002,Abdelqader:2015}. This motivates the use of tidal observables as invariant diagnostics of strong-curvature regions in more general compact-object spacetimes.\\

It should be noted that the likelihood of observing tidal disruption of NSs by supermassive NaSs is dependent on the pulsar or NS population close to galactic centers, in addition to the presence or population of supermassive NaSs. For instance, our Galactic Center has a ``missing pulsar problem" because hardly any pulsars have been discovered there. There may be fewer pulsars close to a galactic center because of the high density of dark matter (DM), which may cause DM particles to gather at the core of the NS and create a tiny BH or NaS, which then accretes and transforms the entire NS into a BH or NaS. Therefore, the absence of NaS may not be indicated by the non-observation or low likelihood of NS disruption close to galactic centers. Therefore, the absence of NSs near galactic centers or the low likelihood of their disruption may indicate the absence of NSs rather than the absence of NaS. \cite{Adarsha}. In Schwarzschild spacetime, the innermost stable circular orbit (ISCO) is located at $ r = 6GM/c^2$. For a test particle in the Schwarzschild geometry, the specific energy at the innermost stable circular orbit (ISCO), located at $r=6GM/c^{2}$, is $E = 2\sqrt{2}/3 = \sqrt{8/9}$. Thus, a particle falling from rest at infinity must radiate an energy $(1 - 2\sqrt{2}/3)m c^{2} \approx 0.057\,m c^{2}$ to reach this orbit, implying an accretion efficiency of $\sim 5.7\%$. These values are determined solely by the orbital dynamics of circular geodesics in the Schwarzschild spacetime and are not specific to any model, which instead provides the detailed radiative structure of the disc.\\

One might naively expect that material approaching $r = 2GM/c^{2}$ could, under ideal circumstances, radiate nearly all of its binding energy, for example if the motion were strongly non–circular and significant dissipation occurred during tidal interactions or infall. In reality, for a Schwarzschild black hole, radiation produced close to the horizon is severely redshifted and most of it is captured rather than escaping to a distant observer. As a result, the radiative output is limited, and the efficiency cannot exceed the standard value determined by the specific energy at the innermost stable circular orbit, $\eta \simeq 5.7\%$.\\

In contrast, for a naked singularity, the absence of an event horizon allows the released binding energy to escape to infinity. In such a case, the theoretical radiative efficiency can, in principle, approach $100\%$ if the infalling matter loses its entire rest–mass energy before reaching the singularity. Here we consider the Joshi-Malafarina-Narayan(JMN) type 1 NaS \cite{psJoshi1}.\\ 

The line element of a general spherically symmetric, static spacetime is given as follows (we discuss it in detail in Appendix \ref{app:AppA_AA_Framework}):   
\begin{equation}
    ds^2 = - f(r)dt^2 + g(r)dr^2 + r^2(d\theta^2 + \sin^2\theta d\phi^2)\,\,. 
    \label{static0}
\end{equation}
\begin{figure}
    \centering
\subfigure[]
{\includegraphics[width=6.0cm]{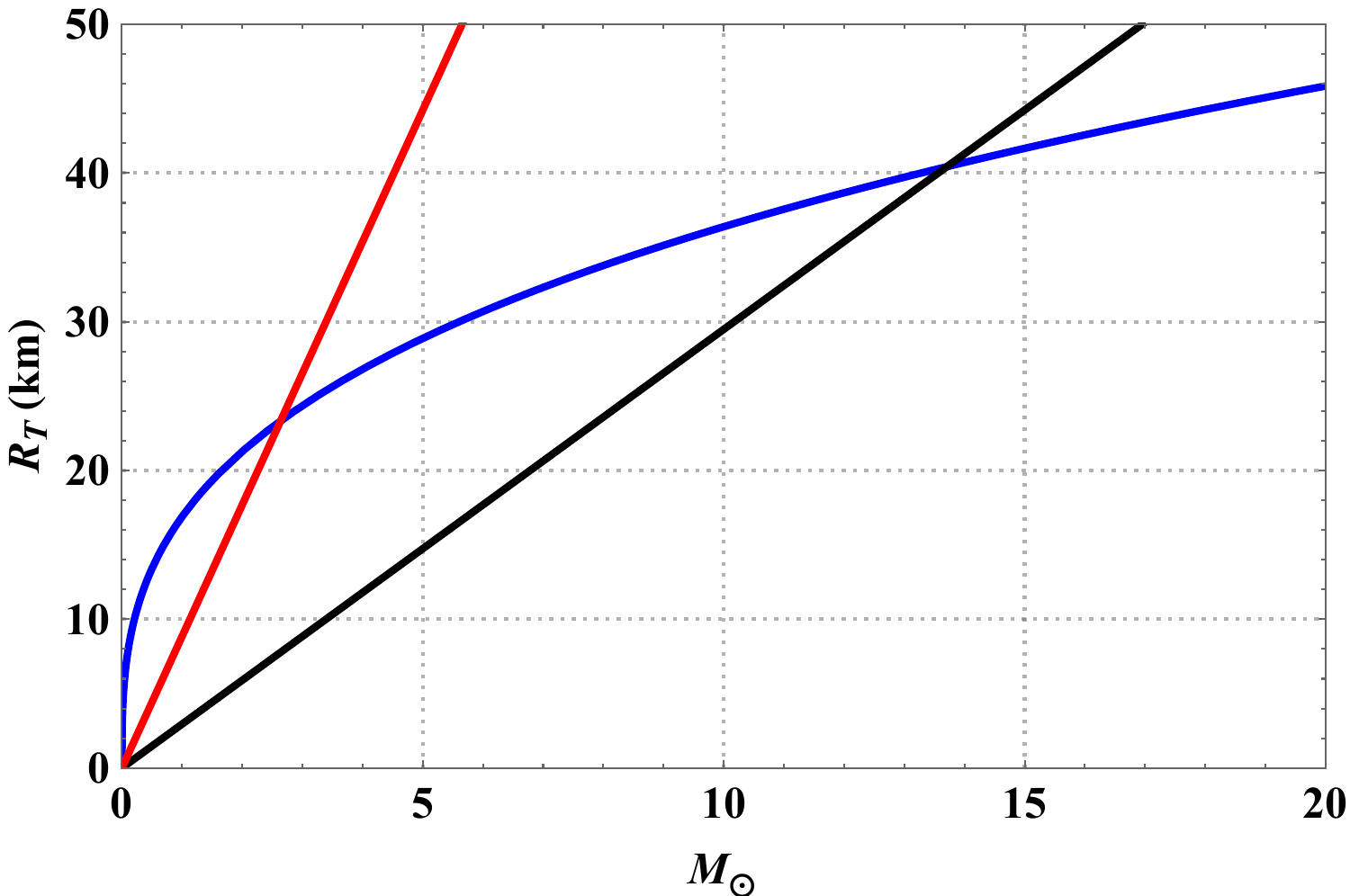}\label{sch}}
\hspace{0.5cm}
\subfigure[]
{\includegraphics[width=6.0cm]{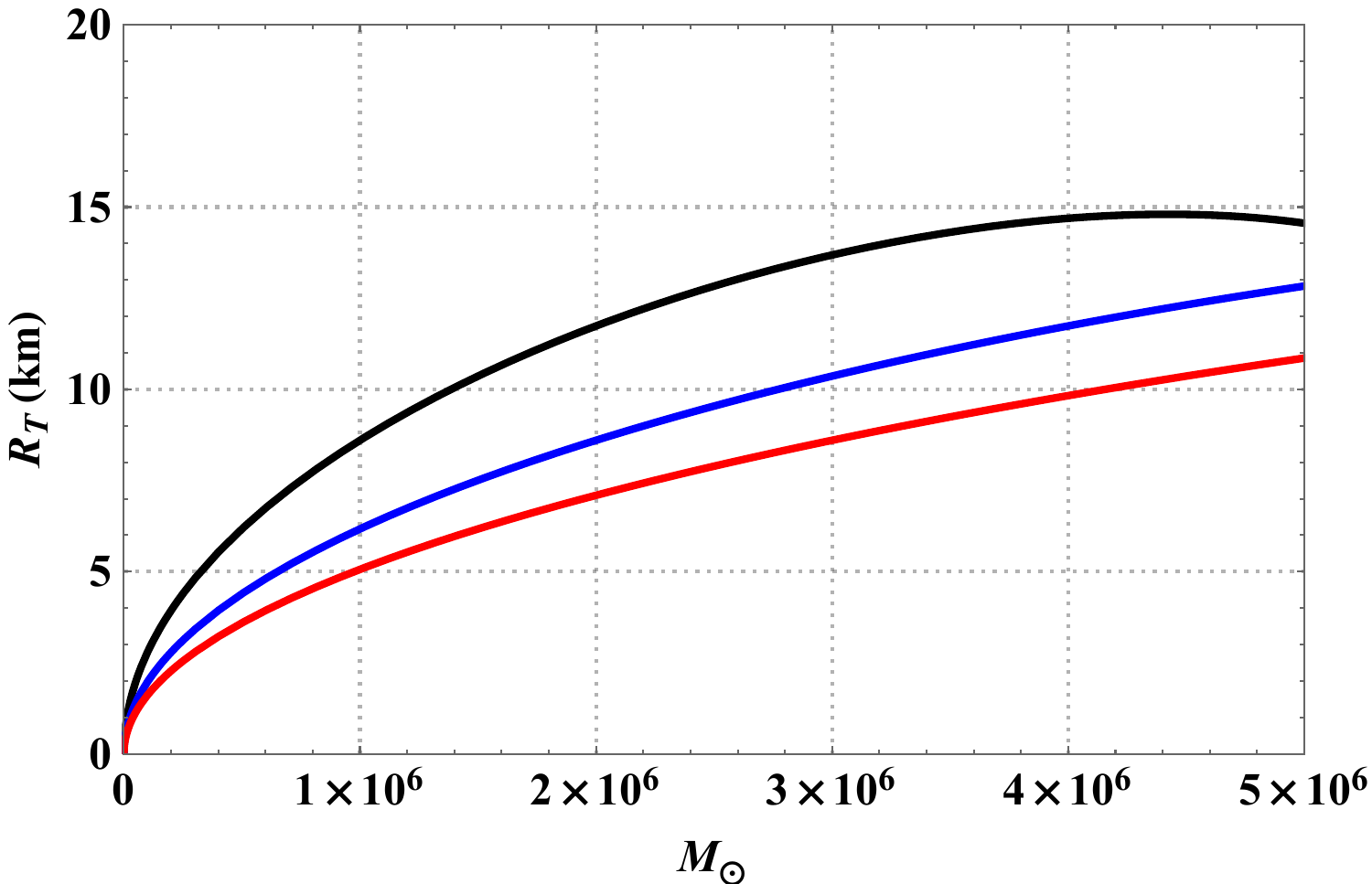}\label{jmn}}
 \caption{In figures, the tidal disruption radius versus the solar mass of the central compact object is shown. Fig.~(\ref{sch}) illustrates how the tidal disruption radius changes with changing the mass of the central Schwarzschild BH. Black, red, and blue lines show the event horizon, innermost stable circular orbit (ISCO), and tidal disruption radius, respectively. Fig.~(\ref{jmn}) demonstrates tidal disruption radius in JMN1 NaS for $M_{0}<2/3$. In the JMN1 case, the tidal disruption radius is directly dependent on the compactness ratio $M_{0}$. We take the NS mass $1.4M\odot$, and the diameter of the NS is $15 \text{km}$. For the JMN1 case, for $R_{b1}$, $R_{b2}$, and $R_{b3}$ are $1.0 \times 10^{-6} Pc$ (black line), $2.0 \times 10^{-6} Pc$ (blue line), $3.0 \times 10^{-6} Pc$ (red line), respectively.}
 \label{figurejacobi}
\end{figure}
For a particle that is freely falling, the angular momentum ($h$) and the energy ($e$) per unit of the particle's rest mass are always conserved, where the angular momentum and energy conservation are the direct consequence of the spherical and temporal symmetry of the above spacetime (\ref{static0}). The conserved quantities $h$ and $e$ of a freely falling particle in static, spherically symmetric spacetime can be written as,
\begin{eqnarray}
    h = r^2\frac{d\phi}{d\tau}\,\, ,\,\,\,
   e =  f(r)\frac{dt}{d\tau}\,\, ,
   \label{congen}
\end{eqnarray}
where $\tau$ is the proper time of the particle and we consider $\theta=\pi/2$. We know that free-falling particles always follow timelike geodesics, for which $v_{\mu}v^{\mu}=-1$, where $v^{\mu}$ is the particle's four-velocity. From this normalization of four-velocity, we can write an effective potential that plays a crucial role in the trajectories of particles in spacetime. For the static and spherically symmetric spacetime, the effective potential can be written as
\begin{equation}
V_{\rm eff} = f(r)\left(1+\frac{h^2}{r^2}\right)
\label{veffgen}\,\, ,
\end{equation}
where the total energy ($E$) can be written as
\begin{equation}
    e^2=f(r)g(r)\left(\frac{dr}{d\tau}\right)^2+V_{\rm eff}(r)\,\, ,
    \label{totalE1}
\end{equation}
For radial timelike geodesics, $\theta$ and $\phi$ are constant. Its angular momentum vanishes, $h=0$, and (\ref{totalE1})) is reduced to 
\begin{equation}
    \dot{r} = \left(\frac{dr}{d\tau}\right) = -\sqrt{\frac{(e^2-f(r))}{f(r)g(r)}}. \label{rdot}
\end{equation}
In the strong-field regime of general relativity, tidal forces provide a physically meaningful and observer-independent probe of spacetime curvature. These effects are most naturally described using the covariant geodesic deviation equation, which encodes the relative acceleration of nearby worldlines directly in terms of the Riemann tensor. The behavior of relativistic tidal stresses has been extensively investigated in black hole and compact-object spacetimes, revealing their sensitivity to both local curvature and global spacetime structure \cite{Marck:1983,Ishii:2005,Chicone:2005}. As such, tidal observables offer a robust framework for diagnosing extreme gravitational environments beyond purely causal or kinematical considerations.
We investigate the effect of tidal forces in a static spherically symmetric (SSS) spacetime. To explore tidal forces in the framework of general relativity, we analyze the geodesic deviation equation:
\begin{equation}
    \frac{D^2 \eta^\mu}{D\tau^2} - R^\mu_{\nu \rho \sigma} v^\nu v^\rho \eta^\sigma = 0, \label{deviationeql1}
\end{equation}
where $ R^\mu_{\nu \rho \sigma} $ and $ v^\nu $ denote the Riemann curvature tensor and the unit tangent vector of the geodesic, respectively, and $ \eta^\mu $ represents the geodesic deviation vector. In the presence of strong energy conditions and non-zero curvature, each point on the test body follows a unique geodesic, leading to stretching and squeezing, known as tidal effects. To quantify these effects, which measures the distance between infinitesimally close geodesics. The tetrad basis related to the freely falling frame is expressed as:
\begin{eqnarray}
\hat{e}^\mu_{\hat{0}} &=& \{\frac{e}{f(r)}, -\left(\frac{e^2-f(r)}{f(r)g(r)}\right)^{1/2}, 0, 0\}, \nonumber \\
\hat{e}^\mu_{\hat{1}} &=& \left\{-\frac{(e^2-f(r))^{1/2}}{f(r)}, \frac{e}{g(r)f(r)}, 0, 0\right\}, \nonumber \\
\hat{e}^\mu_{\hat{2}} &=& \{0, 0, \frac{1}{r}, 0\}, \nonumber \\
\hat{e}^\mu_{\hat{3}} &=& \{0, 0, 0, \frac{1}{r\sin\theta}\}. \label{tetrad1}
\end{eqnarray}
These satisfy the orthonormality condition:
\begin{equation}
    \hat{e}_{\alpha \hat{\mu}} \cdot \hat{e}^\alpha_{\hat{\nu}} = \eta_{\hat{\mu}\hat{\nu}},
\end{equation}
where $ \eta_{\hat{\mu}\hat{\nu}} = \text{diag}(-1, 1, 1, 1) $ represents components of the Minkowski metric. The hat indices denote the tetrad basis, while those without denote the coordinate basis. Additionally, $ \hat{e}^\mu_0 = v^\mu $, implying that the tetrad basis equals the 4-velocity vector of the observer when timelike \cite{zhang2018tidal}. The unit vectors $ \{\hat{e}^\mu_1, \hat{e}^\mu_2, \hat{e}^\mu_3\} $ correspond to orthogonal spatial directions in the observer's frame \cite{chandrasekhar1983mathematical}. Using the tetrad basis from Eq. (\ref{tetrad1}), the geodesic deviation vector (or separation vector) can be expanded as:
\begin{equation}
    \xi^\mu = \hat{e}^\mu_{\hat{\nu}} \xi^{\hat{\nu}}.
\end{equation}
It's important to note that for a fixed temporal component, $ \xi^{\hat{0}} = 0 $ \cite{madan2022tidal}.
Next, we calculate the Riemann curvature tensor with respect to the tetrad basis using the tetrad formalism, given by:
\begin{equation}{\label{riemann tensor}}
    R^{\hat{a}}_{\hat{b}\hat{c}\hat{d}} = R^\mu_{\nu\rho\sigma} \hat{e}^{\hat{a}}_\mu \hat{e}^\nu_{\hat{b}} \hat{e}^\rho_{\hat{c}} \hat{e}^\sigma_{\hat{d}}.
\end{equation}
In the instantaneous rest frame (IFR), Eq.~(\ref{deviationeql1}), can be expressed as
\begin{equation}
\frac{d^2\,\eta^{\hat{\alpha}}}{d\tau^2} = R^{\hat{\alpha}}_{ \hat{0} 
\hat{0} \hat{\gamma}} \, \eta^{\hat{\gamma}}, \label{Rhat1}
\end{equation}
Considering the vectors are parallelly transported along the 
geodesic and exploring the above equations (\ref{tetrad1}) and 
(\ref{Rhat1}), we obtain the relative acceleration between two nearby particles in radial and tangential directions as follows:
\begin{equation}
\frac{d^2 \eta^{\hat{r}}}{d\tau^2}= \frac{ f(r) f'(r) g'(r) + g(r)(f'(r)^2 - 2f(r)f''(r))}{4 f(r)^2  g(r)^2}\eta^{\hat{r}}, \label{tidalg}
\end{equation}
\begin{equation}
\frac{d^2 \eta^{\hat{i}}}{d\tau^2}= \frac{g'(r)f(r)^2 - e^2\left(g(r)f'(r)+ f(r)g'(r)\right)}{2r g(r)^2 f(r)^2} \eta^{\hat{i}}, \label{tidalg1}
\end{equation}
where $i = \theta, \phi$. The above two equations represent the tidal force for free-falling test particles. The extremum of the radial tidal force function can be calculated as follows: 
\begin{equation}
    \frac{d}{dr}\left(\frac{ f(r) f'(r) g'(r) + g(r)(f'(r)^2 - 2f(r)f''(r))}{4 f(r)^2 g(r)^2}\right) = 0.\label{tidalhorizon}
\end{equation}
The maximum positive value of radial tidal force at a fixed radius provides information about the maximum stretching of the matter field within the spacetime.

The spread in specific orbital energies imparted to the stellar material during a tidal encounter primarily reflects the differential gravitational potential across the star evaluated at the tidal radius, under the assumption that the star retains approximate hydrostatic balance prior to disruption (see Lacy et al.\ 1982 \cite{Lacy:1982}). In this picture, fluid elements on the near and far sides of the star gain and lose orbital energy respectively due to the tidal field, producing the characteristic energy distribution that governs fallback rates after disruption.

A significantly broader dispersion of internal energies characterizes the star following the encounter, with some fluid having a positive energy (and thus staying unbound) and some having a negative energy (and hence being bound to the BH). The fluid elements move in Keplerian orbits. Following a Keplerian period T, the bound debris return near the pericenter and are connected to their (negative) energy E by:
\begin{equation}
    E = -\frac{m}{2}\left(\frac{2\pi G M}{T}\right)^{2/3}.\label{energy}
\end{equation}
The basic premise is that the flare is caused by the bound material abruptly accreting onto the SMBH after losing its energy and angular momentum in a period significantly shorter than T after returning to the pericenter. Thus, the mass distribution return to pericenter is essentially the BH's mass accretion rate during the event, from which the brightness can be estimated. Using Eq.(\ref{energy}), we can calculate mass fallback $dm/dT$. Consequently, we have a mass fallback rate:
\begin{equation}
    \frac{dm}{dT} = \frac{dm}{dE}\frac{dE}{dT} = \frac{m(2\pi GM)^{2/3}}{3}\frac{dm}{dE}T^{-5/3}\label{dmdt}
\end{equation}
The fallback rate implied by Eq.~(16) is often associated with the canonical
$\dot{M} \propto t^{-5/3}$ scaling. It is important to note, however, that this
behaviour is strictly asymptotic. While the early-time fallback can deviate
significantly from a pure power law—particularly during the return of the most
bound debris—the late-time behaviour approaches the $t^{-5/3}$ scaling provided
that a nonzero amount of material occupies orbits near $E = 0$, as is evident
from Eq.~(16). This asymptotic behaviour does not require a perfectly flat
$dM/dE$, but follows generically once debris with small specific energies is
present.\\

From an astrophysical perspective, tidal effects play a central role in a variety of high-energy phenomena, most notably tidal disruption events (TDEs), where stars experience extreme relativistic tidal stresses during close encounters with compact objects. The fallback dynamics, luminosity evolution, and observational signatures of TDEs are known to be highly sensitive to the strong-field spacetime geometry and relativistic corrections to stellar motion \cite{Guillochon:2013,Tejeda:2017}. This sensitivity makes tidal observables a promising bridge between theoretical models of strong-field gravity and astrophysical data, particularly in regimes where horizonless compact objects or nonstandard spacetime geometries may be relevant. Deviations from the canonical fallback rate can arise in cases of partial
disruption or when the stellar internal structure plays an important role, as
demonstrated in both analytical and numerical studies \cite{Lacy:1982,Lodato:2009}. These effects primarily influence the
early-time evolution of the fallback rate, while the late-time behaviour remains
consistent with the $t^{-5/3}$ scaling. Observational studies have also tested
this prediction by comparing measured tidal disruption event light curves with
theoretical fallback models, finding broad consistency with the asymptotic
$t^{-5/3}$ behaviour fallowing for significant early-time deviations.\\

The efficient r-process nucleosynthesis observed in GW170817 highlights the crucial role of tidal disruption and mass ejection in heavy-element production. The suppression of comparable r-process signatures in neutron star–black hole mergers indicates that horizon absorption limits ejecta formation, whereas horizonless compact objects may sustain strong tidal interactions capable of enhancing mass ejection and nucleosynthesis, even at high total masses.\\

The plan of the paper is as follows. In Section (\ref{Sec2}), we discuss the tidal disruption radius in a static and spherically symmetric Schwarzschild BH and JMN1 NaS spacetime. In Section (\ref{section3}), we discuss the evolution of flares in the background of BHs and naked singularities. We show a relativistic correction in a standard profile of the light curve $t^{-5/3}$ and the light curve profile for the JMN1 spacetime. Finally, our results are discussed in Section (\ref{sec3}), followed by the conclusion. We take into account $c=1$ the concept of conserved quantity per unit rest mass in the calculation above; we shall restore it for a particular solution later. Throughout the paper, the metric signature $(-,+,+,+)$.

\section{Tidal disruption radius in BHs and  naked singularities}\label{Sec2}
The way stars are destroyed close to supermassive compact objects, which might not be typical black holes but instead have scalar "hair" or even be naked singularities is examined in \cite{Andre:2024bia}. The authors demonstrate how scalar fields can significantly increase tidal forces in the vicinity of black holes. At the same time, flares may be powered by stellar collisions with a dense "gray shell" in the case of a naked singularity. According to their findings, tidal disruption events might provide a means of differentiating between black holes and other, more unusual central objects in galaxies.
The tidal disruption radius is calculated on the principle of the radial tidal force exerted on the test particle, Eq.~(\ref{tidalg})(here we consider a NS with some solar mass), and the gravity on the surface of the test particle is equal. 
However, assuming that the tidal disruption of the star takes place when the self-gravitational attraction on its surface equals the tidal force, which is an elementary but useful approximation for the tidal disruption radius,
\begin{equation}
    \mid{T_{R}}\mid R_{n} = \frac{Gm}{R_{n}^2}. \space \label{roche}
\end{equation}
Equation~(17) is adopted from previous studies of tidal disruption in black hole
and naked singularity spacetimes \cite{Andre:2024bia} and should
be understood as a heuristic criterion indicating the onset of significant tidal
deformation, rather than as a fully relativistic tidal radius. In a proper
relativistic treatment, tidal effects are characterized by the eigenvalues of
the tidal tensor constructed from the Riemann curvature projected onto the local
orthonormal frame of the neutron star, and quantities such as the stellar radius
cannot be interpreted as coordinate distances. Fully relativistic analyses have shown that the tidal disruption threshold
depends sensitively on the neutron star compactness, equation of state, and the
spacetime curvature of the central object \cite{Thorne:1987,Hinderer:2008,Beloborodov:1992,Kesden:2012,Stone:2013}. Our use of Eq.~(17) is therefore
intended to provide qualitative insight into the relative strength of tidal
effects in naked singularity spacetimes compared to black holes, rather than a
precise or invariant disruption criterion. Here, the radial tidal force component is, 
\begin{equation}
    T_R = \frac{1}{\eta^{\hat{r}}} \frac{d^2 \eta^{\hat{r}}}{d\tau^2},
\end{equation}
where $R_{n}$ is the distance between the gravitational center to the surface of the test particle (here we consider a NS), and $m$ is the mass of the test particle. This tidal disruption radius sometimes it is called as Roche limit or Roche radius. As we considered, the test particle has a much smaller mass than the central compact object. It is emphasized that in the Newtonian limit, surface gravity on the test particle should be $Gm/R_{n}^{2}$. This Roche limit can be divided into two parts: the radial component of the Roche limit radius $(R^{r}_{t})$ and the angular component of the Roche limit radius $(R^{\theta}_{t})$.

\subsection{Roche limit in Schwarzschild spacetime}
When a NS approaches a supermassive BH that is inside its tidal disruption radius, it tidally disrupts the star. For radially freely falling test particles in the background of a Schwarzschild BH, the radial tidal force is:
\begin{eqnarray}
    T_{R} = \frac{2GM}{r^3},\label{tidal}
\end{eqnarray}
where $M$ is the mass of the BH.
Now putting Eq.~(\ref{tidal}) into Eq.~(\ref{roche}) gives the Roche limit,
\begin{eqnarray}
  R^{r}_{t} = R_{T} = R_{n} \left(\frac{2M}{m}\right)^{1/3}.
\end{eqnarray}
The majority of neutron stars are near 1.4 solar masses, with a typical mass of 1.1 to 2.3 times that of our Sun \cite{Bhattacharyya:2017tos,Ozel:2016oaf}. They are believed to collapse into black holes above this range, and they cannot form below it.
For the typical NS case, $m=1.4M_{\odot}$, $R_{n} = 15\text{km}$. For our Milkey Way galaxy, the central BH mass of Sgr A* is around $4.3 \times 10^6 M_{\odot}$. Therefore, the tidal disruption radius is $R_{T} = 1831.43 \text{km}$. However, the event horizon of Sgr A* is at $1.27 \times 10^7 \text{km}$. Hence, tidal disruption of NSs in Sgr A* is never visible from the Earth.

For the given profile of a NS, a tidal disruption radius forms at the event horizon when the mass of the central BH is $13.7 M_{\odot}$ as shown in Fig.~(\ref{sch}) (crossing lines of black and blue). Similarly, as shown in Fig.~(\ref{sch}), tidal disruption radius forms in the BH at ISCO when the mass of the BH is $2.64 M_{\odot}$ (crossing lines of red and blue). As a result, a high-energy collision within the $6M$ range caused by NS tidal disruption is not observable to observers on Earth. The innermost turning point in Schwarzschild spacetime is $4M$; however, due to tidal disruption from an almost radial falling NS, the average angular momentum is always much smaller, and all matter eventually falls into the BH. As a result, observers on Earth can see tidal disruption within the $6M$ for a brief period of time. Because the collision period is small, energy flux coming out is comparably smaller than the collision above the $6M$ radius. A small mass BH, such as $2M_{\odot}$, forms a close pair with a NS. As a result,  the tidal disruption of a NS within $6GM/c^2$ is very different from that of a single-body system.


\subsection{Roche limit in JMN1 NaS spacetime}
The line element of the JMN1 spacetime,
\begin{eqnarray}
 ds^2=-(1-M_0) \left(\frac{r}{R_b}\right)^\frac{M_0}{1-M_0}dt^2 + \frac{dr^2}{1-M_0} + r^2d\Omega^2\,\,,  
\label{JMN-1metric}
\end{eqnarray}
where the dimensionless parameter $M_0$ should be $0<M_0<1$ and $R_b$ is boundary radius of JMN1 spacetime that matches with the external Schwarzschild spacetime. It is demonstrated in \cite{psJoshi1} that JMN1 spacetime can form as an endstate of the gravitational collapse of an anisotropic matter fluid in asymptotic time. At the coordinate center, this spacetime has a null singularity for $M_{0}>2/3$ and a timelike singularity when $M_{0}<2/3$ \cite{Bambhaniya:2021jum}. For the JMN1 case, total mass within the radius $R_{b}$ is given by,
\begin{equation}
    M_{0} = \frac{2GM}{c^2 R_{b}}.\label{Mpara1}
\end{equation}
For radially freely falling test particles in the background of a JMN1 NaS spacetime, the radial tidal force is,
\begin{equation}
   T_{R} = \frac{c^2 M_{0}\left(2 - 3M_{0}\right)}{4 \left(1 - M_{0}\right) r^2},\label{tidaljmnr}
\end{equation}
where $G$ is gravitational constant and $c$ is velocity of light.
Now putting Eq.~(\ref{Mpara1}) and Eq.~(\ref{tidaljmnr}) in Eq.~(\ref{roche}) gives,
\begin{equation}
     R_{T} = R_{n} \sqrt{\frac{\left(1 - \frac{3GM}{c^2 R_{b}}\right) M R_{n}}{\left(1 - \frac{2GM}{c^2 R_{b}}\right) m R_{b}}}.\label{rochejmn}
\end{equation}
We may choose $m=1.4M_{\odot}$, $R_{n} = 15\text{km}$ for a typical NS case. For our Milkey Way galaxy, the central mass of Sgr A* is around, $4.3 \times 10^6 M_{\odot}$, and for JMN1 spacetime with a finite boundary ($R_{b}$) with $3.085\times10^7$ kilometers. Hence, the parameter value that determines the mass, $M_{0} = 0.41$, the tidal disruption radius for an ideal NS is around $14.791$ kilometers. Comparatively speaking, the tidal disruption radius in the JMN1 spacetime is around 124 times smaller than the tidal disruption radius of a Schwarzschild BH.

Fig.~(\ref{jmn}) demonstrates tidal disruption radius in JMN1 NaS for $M_{0}<2/3$. As shown in Eq.~(\ref{rochejmn}), a real positive value of $T_{R}$ is when $M_{0} < 2/3$, which implies that tidal disruption is because of the stretching forces while $M_{0} > 2/3$ compressive forces are present. Tidal deformation is also closely related to the Jacobi fields that are fully explained in Appendix \ref{appB}.

\section{relativistic timescale for bound and unbound debris}\label{section3}
The evolution of flares with time near a supermassive compact object or the merger of two compact binaries arises due to the disruption of the compact object. Here, we analytically show the difference between naked singularities and BHs, relating the light curve to the internal density distribution of the star. As it has been previously shown \cite{Lodato:2009}, the standard light curve is proportional to $t^{-5/3}$ only holds at late times. Unbound debris always carries larger mechanical energies than the rest mass energy.

\begin{figure*}
\centering
\includegraphics[width=8.0cm]{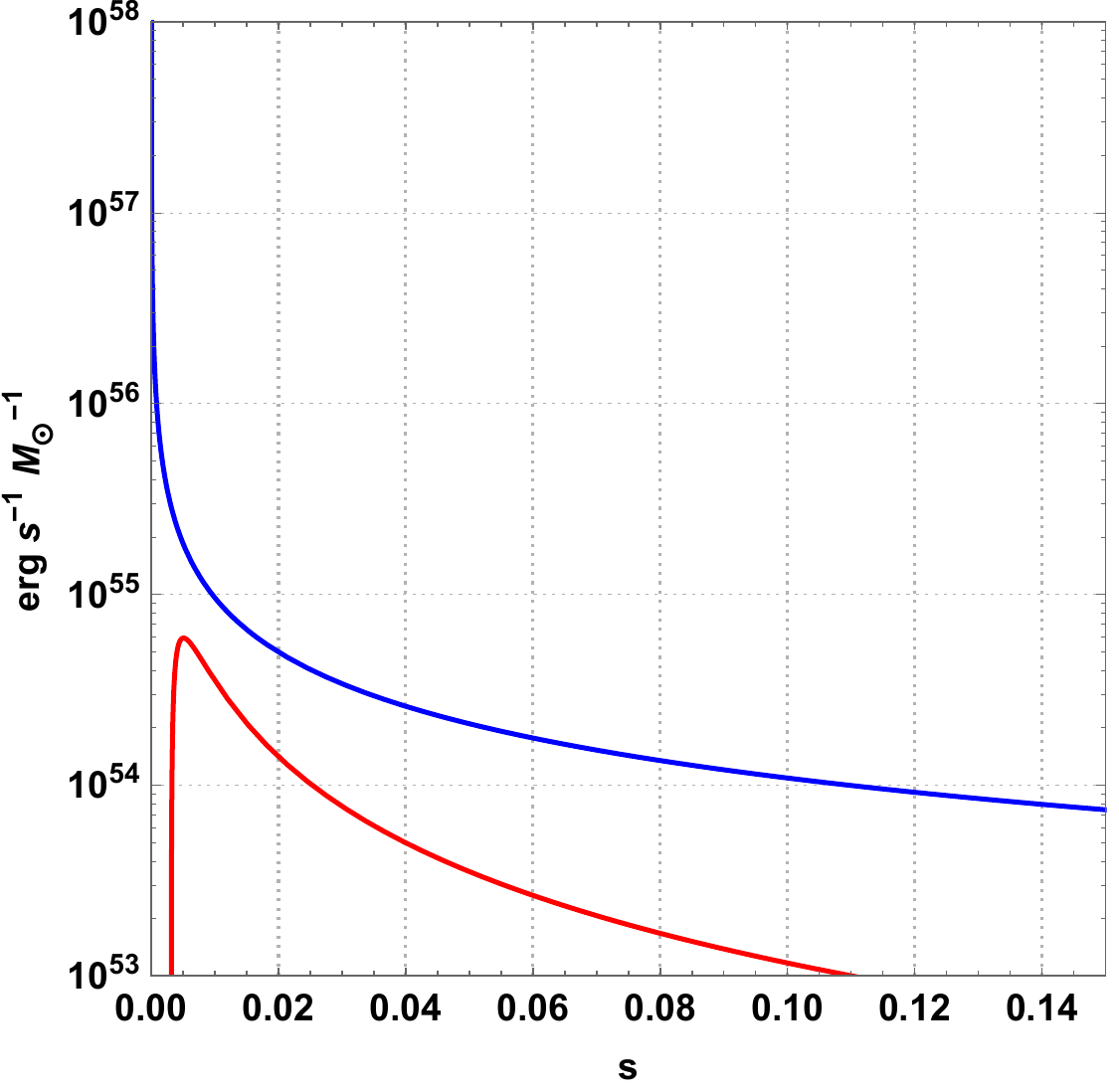}
\caption{The figure shows a light curve in the BH and JMN1 NaS. The red and blue curves show the light curve in the BH and JMN1 NaS, respectively. We considered both have $10M_\odot$ and in JMN1 boundary radius $R_{b} = 10GM/c^2$.}
 \label{fig2}
\end{figure*}

During the disruption, the momentum and energy distribution within the star will decide the free particles' path. The star's core density is higher than the surface density due to inhomogeneity in the matter distribution. Hence, the core's heavy elements transfer energy and momentum to the lighter elements at the star's outer surface. Therefore, matter at the outer surface is expelled first, and then core disruption occurs in the second stage.

Tidal disruption process could be divided into four major parts:
\begin{enumerate}
    \item All matter expelled out of the stars (NSs) has high positive energy ($e>>1$) and large angular momentum on the tidal disruption radius. The debris of these types of stars is unbound and has high-energy, bright flares visible for a short period.
    \item Debris of some finite negative energy stars ( $e<1$) with angular momentum $h>4GM/c$, are bound in the system, and they lose $h$ and $e$ very slowly. They eventually create an accretion disk around the compact objects. These are also responsible for the quenching of the star in the relatively quiescent galaxies (environmental quenching or tidal stripping).
    \item Small negative energy ( $e<<1$) and angular momentum ($h<<2\sqrt{3}GM/c$) debris, or we can say in the BH case that tidal disruption within $6GM/c^2$, debris always terminates into the BH.
    \item In the intermediate state, I) some debris is bound and some falls into the compact object, and II) some debris is bound and some is unbound. As already stated, Rees \cite{Rees:1988} investigated the II case.
\end{enumerate}
The tidal disruption of a NS is in the strong-gravity region. Hence, the Keplerian law could not be valid (Eq.(\ref{dmdt})). High-energy collisions are the reason behind high-energy gamma-ray bursts (GRB). The initial collision is because of the circular orbit, which takes less time than the elliptical orbit. The proper time for a circular orbit is given as follows:
\begin{equation}
    \tau = \frac{2\pi r^{3/2}}{\sqrt{G M}}\sqrt{1-\frac{3G M}{c^2 r}}
\end{equation}
For the BH (static) case, the closest circular orbit is at $6M$, therefore, the smallest collision period can be given as,
\begin{equation}
   \tau = \frac{12\sqrt{3}\pi G M}{c^3}
\end{equation}
Energy in circular orbit for 1st-order correction can be given as follows:
\begin{equation}
    E \approx -\frac{GMm}{2r}+\frac{3G^{2}M^{2} m}{8c^{2}r^{2}}.
\end{equation}
Therefore, the variation of energy distribution over proper time with relativistic correction in the Schwarzschild BH is given as follows:
\begin{equation}
   \frac{dE}{d\tau} = \frac{m(2\pi GM)^{2/3}}{3 \tau^{5/3}} - \frac{7m(2\pi GM)^{4/3}}{6c^2 \tau^{7/3}} - \frac{15m(2\pi GM)^{2}}{8c^4 \tau^{3}}+ \mathcal{O}(\tau^{-11/3}).
\end{equation}
As it is shown in the strong gravity correction order $1/c^2$, $dE/d\tau>0$. From Fig.(\ref{sch}),  we can conclude that tidal disruption within the ISCO to the event horizon is possible within the stellar mass mergers of binary BH and NS ($1.5 M_{\odot}$ to $7.3 M_{\odot}$). In the Schwarzschild case, the innermost unstable circular orbit (IUCO) is $4M$. Hence, in between the $4M$ to $6M$ unstable circular orbit can be possible. If the tidal disruption radius is within this range, the tidal disruption radius is the periapsis of the NS. If a NS approaches a BH so closely that it is affected by tidal distortions and even disruption, it cannot be considered a ``point mass" interacting with the BH. Such effects become significant when the `tidal radius' is as small as the pericenter, $R_{T} <r_{min}$. Therefore, complete disruption requires $\beta = R_{T}/ r_{min} <1$.

For JMN-1 NaS proper time for a circular orbit is given as follows:
\begin{equation}
    \tau = \frac{2\pi r \sqrt{R_{b}}}{\sqrt{G M}}\sqrt{1-\frac{3G M}{c^2 R_{b}}}
\end{equation}
We can analyze that $R_{b}>3GM/c^2$ from the above equation. Circular orbits for $M_{0}>2/3$ will never form. The energy radiation rate for the NaS case is given as follows :
\begin{equation}
    \frac{dE}{d\tau} = \left(\frac{GMm\left(\frac{GMc^2}{c^2 R_{b}-3GM}\right)^{\frac{GM}{2c^2 R_{b}-4GM}}}{R_{b} \left(2\pi R_{b}\right)^{\frac{GM}{c^2 R_{b}-2G M}}\sqrt{1-\frac{3GM}{c^2 R_{b}}}}\right) \tau^{-\frac{(c^2 R_{b}-3GM)}{(c^2 R_{b}-2G M)}}.
\end{equation}
Suppose we assume that the angular momentum loss is significantly smaller than the energy loss. In that case, we can conclude that angular momentum remains constant over time, while the energy spread is significant. As a result, in this scenario, the maximum spread of energy can be $\Delta E = E_{max} - E_{min}$. Here, $E_{max}$ and $E_{min}$ are energies in unstable and stable circular orbits.

In Fig.(\ref{fig2}), we show the light curve in a Schwarzschild BH and a JMN1 NaS. As expected in the NaS case, the energy emission in the strong gravity region is much higher. In late time emission in NaS is more than 10 time of a BH. If we consider the tidal disruption of a typical NS in a $10M_\odot$ BH. It will emit a first maximum energy flare that is approximately $5.8\times10^{54} erg M_\odot^{-1} s^{-1}$ at the duration $5ms$. In JMN1 NaS, at the time of $5ms$ energy emission is around $1.8\times10^{55} erg M_\odot^{-1} s^{-1}$.\\

\subsection{Bound debris}
\subsubsection{Schwarzschild Black Hole}
In the bound debris, the energy of the debris is negative $E<0$. The matter field is bound between the two turning points, the pericenter and the apocenter. The radial escape velocity in the Schwarzschild BH is as follows:
\begin{equation}
     V_{e} = \sqrt{\frac{2GM}{r} - \frac{L^2}{m^2 r^2}\left(1-\frac{2GM}{c^2 r}\right)}.
\end{equation}
Where, $L = m h$. The velocity of the debris is not larger than the escape velocity, $V<V_{e}$. The disruption impulse at the pericenter occurs immediately and affects the star's spin to some extent but does not alter the kinetic energy of the fluid elements. Angular velocity of the star at pericenter is given as follows: 
\begin{equation}
    \Omega_{p} = \frac{d\phi}{dt} = \sqrt{\frac{2GM}{r_{p}^3}\left(1-\frac{2GM}{c^2 r_{p}}\right)},
\end{equation}
where the $r_{p}$ is the closest approach of debris (for a parabolic orbit with $E=0$) and $t$ is coordinate time. For bound debris, $\Omega < \Omega_{p}$. Minimum energy, escape velocity, and angular velocity at the ISCO are given as follows:
\begin{equation}
    E_{min} = -0.05719 m c^2, \hspace{0.3cm} \frac{dr}{d\tau} = \frac{c}{3}, \hspace{0.3cm}  \Omega_{p} = \frac{c^3}{18GM}.
\end{equation}
Suppose a stable circular orbit loses its energy and angular momentum in such a manner that it eventually remains in the circular orbit. In that case, the orbital radius will always be decreasing. However, if an unstable circular orbit within the range of $4M$ to less than $6M$ loses its energy or angular momentum, its orbital radius increases instead of decreasing. Therefore, collisions between the stable and unstable circular orbits are possible near the $6GM/c^2$. During the tidal disruption event of sufficiently low angular momentum and energy particles, always create a high luminosity glow around the ISCO. Hence, high-energy photosphere-like structures form near ISCO.
\subsubsection{JMN1 Naked Singularity}
In the case of the NaS of JMN1, the energy at the space-time boundary $R_{b}$ is always negative. Hence, particles have always plunged into and bound to orbits. Therefore, the energy needed to escape the particle from the boundary of JMN1 spacetime is given as follows:
\begin{equation}
    E_{b} = \left(\sqrt{1-\frac{2GM}{c^2 R_{b}}} \left(1+\frac{L^2}{c^2 m^2 R_{b}^2}\right)^{1/2}-1\right)m c^2
\end{equation}
If the energy of the debris is $E$, $E > E_{b}$, the debris crosses the boundary radius $R_{b}$ and spreads in the Schwarzschild spacetime, has an energy range $E_{b}<E<0$. However, the debris is bound in Schwarzschild spacetime. Any particle with non-zero angular momentum (even arbitrarily close to zero) is always bound within spacetime for energy $E<E_{b}$.
The escape velocity of particles in proper time for debris in JMN1 spacetime is given as follows:
\begin{equation}
    V_{e} = c\sqrt{\left(\frac{R_{b}}{r}\right)^{\frac{M_{0}}{(1-M_{0})}}-(1-M_{0})\left(1+ \frac{L^2}{m^2 c^2 r^2}\right)}.
\end{equation}
For the bound debris velocity is not larger than the escape velocity, $V_{e}>V$. Similarly, angular velocity in the JMN1 spacetime at the turning point of debris is given as follows:
\begin{equation}
    \Omega_{p} = \frac{c \sqrt{1-M_{0}}}{r_{p}}\left(\frac{r_{p}}{R_{b}}\right)^{\frac{M_{0}}{2(1-M_{0})}} \sqrt{1-(1-M_{0})\left(\frac{r_{p}}{R_{b}}\right)^{\frac{M_{0}}{1-M_{0}}}}
\end{equation}
For bound debris, $\Omega < \Omega_{p}$. In the Schwarzschild case $\Omega_{p}=0$ at $r=2GM/c^2$, but in the JMN1 NaS it is at $r_{p} \to 0$. As it is known in the Schwarzschild case, the smallest inner turning point is $4GM/c^2$. Therefore, the smallest $\Omega_{p}=c^3/8GM$.\\
\subsection{Unbound debris}
Even with small angular momentum, unbound debris can be possible near the singularity in some naked singularities. On the other hand, the closest unbound debris in the Schwarzschild BH is $4GM/c^2$. If in the Schwarzschild case, the tidal disruption radius is within the $4GM/c^2 < r< 6GM/c^2$, the angular momentum for unbound debris follows the inequality:
\begin{equation*}
    \frac{4GMm}{c} < L < \frac{3\sqrt{2}GMm}{c}.
\end{equation*}
Escape velocity in unbound case ($E>0$) must follows, $V_{e}<dr/d\tau$. Unbound debris produces higher energy photons than bound debris. Initial flares in unbound debris can emit photons with energies of roughly $1.2 MeV$, which can possibly decay into electrons and positrons. This process is only observable during the tidal disruption of unbound debris. A high kinetic energy of a neutron particle decays into a proton and an electron, 
\begin{equation*}
    n \to p^{+} + e^{-} + \gamma
\end{equation*}
If a photon has high enough energy, it can decay further into a positron and an electron,
\begin{equation*}
    \gamma \to e^{+} + e^{-}
\end{equation*}
These high-energy processes do not last for more than a few microseconds. However, in the NaS case, small angular momentum can cause the formation of a circular orbit.
Unbound debris with a turning point that verges near the singularity, which is smaller than $6GM/c^2$ in the case of a BH. The contribution of these materials to jet formation is significantly greater.

In NaS-NS binaries, the tidal disruption of the NS produces electromagnetic counterparts that span the entire electromagnetic spectrum, including gamma-ray bursts (GRBs) and infrared kilonova signals. A portion of the solar mass escapes into space as neutron-rich matter due to the tidal disruption of NSs, which causes them to carry out rapid neutron capture nucleosynthesis and produce heavy metals like gold and platinum. A fast-expanding, somewhat isotropic thermal transient known as a ``kilonova" is powered by the radioactive decay of unstable nuclei produced by the fission and fusion processes in high kinetic energy particles (the matter that is expelled)\cite{Metzger:2019zeh}. \\

\section{Discussion and Conclusion}{\label{sec3}}
The causal structure of naked singularities differs significantly from that of BHs. The existence of naked singularities is frequently contested in the scientific community. Important information regarding naked singularities and BHs can be obtained from observational data. The following information can be obtained from TDEs and matter field motion: 

\begin{itemize}
    \item We investigated a fascinating issue, ``Can tidal disruption of a NS be observable to an asymptotic observer in the situation of a Schwarzschild BH?" If the core BH mass is bigger than approximately $14M_{\odot}$, the tidal disruption of a NS will not be observable to distant observers. Therefore, the tidal disruption of a NS by a supermassive BH will never be observable to the asymptotic observer. As a result, if we assume that the heart of our Milky Way galaxy is a BH, we will be unable to see the tidal disruption of NSs in our galaxy.

    \item In case of tidal disruption within the ISCO, the mass of the central BH should be at maximum $2.64 M_{\odot}$. In the NS-NS merger, a rapid neutron-capture process (r-process) occurs, which causes the formation of heavy elements. However, in the BH case, there are fewer chances to observe the r-process. Only in the small mass (stellar BH $20M_{\odot} <M$) BH-NS binary, the r-process is observable. There is no mass limit in the NaS case; therefore, r-processes are always visible in naked singularities.

    \item Broderick and Salehi \cite{Broderick:2024vjp} have just published an intriguing research study that demonstrates the strong probability of a photosphere in a NaS due to the inner turning point. Here, we demonstrate that a photosphere-like structure will also arise at $6GM/c^2$ (a blazing surface) in BH cases. Observationally, the two bright rings that must be observed are the photon ring and the ISCO radii.

    \item In the late time, the light curve in BHs is proportional to $t^{-5/3}$. However, in the NaS, particularly JMN1, spacetime is dependent on the compactness ratio, which is given as,
\begin{equation*}
    \frac{dE}{dt} \propto t^{-5/3}, \hspace{1.0cm} \frac{dE}{dt} \propto t^{-\frac{2-3M_{0}}{2(1-M_{0})}}
\end{equation*}
In conclusion, in a finite boundary spacetime, the light curve is dependent on the compactness ratio $M_{0}$. An interesting result we find in our analysis is that, in the JMN1 spacetime, for the range $M_{0} \in (0,2/3)$, the corresponding power-law index $n$ lies within the interval $(1,0)$. This range aligns well with the best-fit power-law models ($n \in (0,1)$) obtained from the full X-ray light curves of known and candidate X-ray TDEs (Tidal Disruption Events), such as NGC 247, 2J049, OGLE16aa, and PTF-10iya. This suggests that the JMN1 spacetime could provide a viable and promising model for describing such phenomena.
\end{itemize}

\section{Summary}
In this paper, we investigated what happens to a neutron star (NS) as it gets closer to a naked singularity (NaS) or a black hole (BH). We demonstrated that the NS is nearly completely ingested by BHs above a specific mass threshold, leaving behind a negligible detectable signature. On the other hand, for both stellar-mass and supermassive configurations, the tidal disruption of the NS can be observed from a distance in the NaS case due to the lack of an event horizon.

The potential for bright kilonova-like transients to be triggered from NS disruption in supermassive compact objects without horizons is a novel aspect of this study. Depending on how much of the disrupted mass escapes, such events could yield significant r-process material and possibly enrich the universe with more heavy elements than typical NS--NS mergers. The observational signatures of these events are significantly shaped by the relativistic timescales for both bound and unbound debris, which we also discussed. Crucially, the two scenarios have different expected light curves: for BHs, it rises to a maximum and then decays, whereas for NaSs, because of the lack of an event horizon, it keeps diverging.

Our analysis emphasizes the astrophysical significance of NS disruptions as possible contributors to heavy-element nucleosynthesis and as probes of strong-gravity regimes. These findings also imply that upcoming multi-messenger observations of unusual transients might provide new methods for differentiating BHs from horizonless compact objects.

\acknowledgments

SB acknowledges financial support by the Fulbright-Nehru Academic \& Professional Excellence Award (Research), sponsored by the U.S. Department of State and the United States-India Educational Foundation (grant number: 3062/F-N APE/2024; program number: G-1-00005).

\appendix

\setcounter{section}{1}
\appendix


\section{Geometrical structure of static Spacetimes}\label{app:AppA_AA_Framework}
A spacetime is said to be spherically symmetric and static if its isometry group contains a subgroup that is isomorphic to the rotation group SO(3) and this group's orbits are 2-spheres that is spacetime does not change over time. For the sake of simplicity, here considered $S^2$ symmetry. The line element is given as follows:   
\begin{equation}
    ds^2 = - f(r)dt^2 + g(r)dr^2 + h(r)(d\theta^2 + \sin^2\theta d\phi^2)\,\,. 
    \label{static1}
\end{equation}
In spherical symmetry, $h(r)$ is such that the function is regular and continuous in the domain of $0<r<\infty$. In our case minimum of function $h(r = 0) = 0$. 
Curvature scalar for spacetime $R_{\alpha \beta \gamma \delta} R^{\alpha \beta \gamma \delta}$ is also dependent on the $h(r)$. In conformal geometry -
\begin{equation}
    ds^2 = \Omega(r)^2 \widetilde{g_{\mu \nu}} dx^{\mu} dx^{\nu},
\end{equation}
where, $g_{\mu \nu} = \Omega(r)^2 \widetilde{g_{\mu \nu}}$, here $\Omega(r)^2$ has functional freedom to choose a function such that a non-singular spacetime can be singular or vice-versa. So to avoid such things and for simplicity, we considered $h(r) = r^2$. Hence, the line element of spacetime can be written as, 
\begin{equation}
    ds^2 = - f(r)dt^2 + g(r)dr^2 + r^2(d\theta^2 + \sin^2\theta d\phi^2)\,\,,
    \label{static3}
\end{equation}
where $f(r)$ and $g(r)$ are only functions of $r$ and the azimuthal part of the spacetime exhibits spherical symmetry. The curvature of spacetime is caused by the distribution of energy or mass. As a result, understanding the general theory of relativity is the distribution of matter. In this section, we revisited Wald's book calculation in detail and a systematic way to analyze the matter distribution. The metric is a function of pressure and density. From the solution of the Einstein field equation, the Energy Momentum Tensor $-T^t _t = \rho$, $T^r _r = p_r$ and $T^\theta _\theta = T^\phi _\phi = p_\theta = p_\phi$, and the geometry can be written as:
\begin{equation}
    \rho = \frac{-g(r) + g(r)^2 + r g(r)^{'}}{\kappa r^2 g(r)^2} , \label{density}
\end{equation}
\begin{equation}
     p_{r} = \frac{ f(r) - f(r) g(r) + r f^{'}(r)}{r^2 f(r) g(r)} \label{pressure}
\end{equation}
where, $\kappa = 8\pi = 1$. By solving Eq.~(\ref{density}) and Eq.~(\ref{pressure}), 
\begin{equation}
    g(r) = \left(1-\frac{ \mathcal{J}(r)}{r}\right)^{-1} ,\,\,\,\,\,\,\,\,\,\, f(r) = e^{\int \left(\frac{( P_{r} r^3 + \mathcal{J}(r)))}{r(r - \mathcal{J}(r))} \right) dr}  \label{g}
\end{equation}
where, $\mathcal{J}(r)$ is the Mass function and $P_{r}$ is the radial pressure in spherically symmetric, static spacetime. Where,
\begin{equation}
    \mathcal{J}(r) = 2 \int \rho(r) r^2 dr.\label{Massf}
\end{equation}
Einstein field solution for Eq.~(\ref{g}) provides a tangential pressure that is dependent on the mass function and radial pressure,
\begin{equation}
    p_{\theta} = \frac{r^5 p_{r}^2 + \mathcal{J}(r) \mathcal{J}^{'}(r) + r^2 p_{r}(r(4 + \mathcal{J}^{'}(r)) - 3\mathcal{J}(r)) + 2r^3 (r - \mathcal{J}(r))p^{'}_{r})}{4 r^2 (r - \mathcal{J}(r))}.
\end{equation}
Using the metric components (\ref{g}), it is difficult to determine such a mass function and radial pressure that yield isotropic pressure; the only way is to solve the differential equation $p_{r} - p_{\theta} = 0$. Solving this differential equation is a challenging task. The conservation property of the energy-momentum tensor. One way to express the fundamental laws of conservation of energy and momentum is by saying:
\begin{equation}
    \nabla_{\nu}T^{\mu\nu} = T_{;\nu}^{\mu\nu} = 0 \label{tmunu}
\end{equation}
where, $T^{\mu\nu}$ is the energy-momentum tensor. Using equation (\ref{tmunu}),  
\begin{equation}
    - P_{r}^{'} = (P_{r}+\rho)\Gamma_{01}^{0}+2\Gamma_{21}^{2}(P_{r} - P_{\theta}) \label{CDEMT}
\end{equation}
Now, by solving the (\ref{CDEMT}) line element can be written as -

\begin{eqnarray}
ds^2 = -e^{\Phi(r)} dt^2+\frac{1}{\left(1-\frac{\mathcal{J}(r)}{r}\right)}dr^2 + r^2d\Omega^2\,\, , \label{genmetric}
\end{eqnarray}

where
\begin{eqnarray}
     \Phi(r) =  \int \left(\frac{( 4(P_{\theta}-P_{r}) - 2r P_{r}^{'})}{r(P_{r} + \rho)} \right) dr\\ = \int \left(\frac{( P_{r} r^3 + \mathcal{J}(r)))}{r(r - \mathcal{J}(r))} \right) dr
\end{eqnarray}
For isotropic pressure that is $P_{r} = P_{\theta} =P_{\phi}=P$ and with equation of state $\omega=P/ \rho$, 
\begin{equation}
    P^2 r^3 + \mathcal{J}(r) P + \frac{2\omega r}{(1+\omega)}P^{'} (r-\mathcal{J}(r))=0
\end{equation}
where,
\begin{equation}
    \mathcal{J}(r) = \frac{1}{\omega}\int P(r) r^2 dr.
\end{equation}
The prime denotes the differentation over the radial distance. If the density distribution and radial pressure are known in the spherically symmetric spacetime, then the metric of the spacetime; for example, if the radial density and pressure profile of the Sun are known, then we can easily derive the spacetime inside the sun. For asymptotically Minkowskian spacetime, we take the condition in which the Kretsmann scalar vanishes at an infinite distance, 
\begin{equation}
    \displaystyle{\lim_{r \to \infty}}(R_{\alpha \beta \gamma \delta} R^{\alpha \beta \gamma \delta}) = 0.\label{vanish}
\end{equation}
Hence the metric components of spacetime resemble with Minkowskian metric in the asymptotic regions,
\begin{equation}
    \displaystyle{\lim_{r \to \infty}} f(r) = 1,\,\,\,\,\,\  \displaystyle{\lim_{r \to \infty}} g(r) = 1.
\end{equation}
It is important to know the energy condition of the spacetime. In general relativity, energy conditions provide information about positive mass and attractive gravity. Now using Eq.~(\ref{Massf}) and Eq.~(\ref{vanish}),
\begin{equation}
    \lim_{r\to \infty} \mathcal{J}(r) = 2 M,
\end{equation}
$M$ is the total mass of the asymptotically Minkowskian spacetime. Suppose $(\mathcal{M}, g)$ is a spacetime satisfying all energy conditions. In that case, the maximal Cauchy developments are future inextendible as a suitably regular Lorentzian manifold for generic asymptotically flat initial data. Spacetime is the exact solution of the Einstein field equation when it must satisfy strong, weak, dominant, null, and null dominant energy conditions. Therefore, to satisfy energy conditions, we require,
\begin{equation}
     \mathcal{J}(r) \geq 0,\,\,\,\,\,\
   \mathcal{J'}(r) \geq  0,\space
\end{equation}
and it can be easily verified that all these necessary conditions are fulfilled for the given spacetime. 
In the vacuum solution of the Einstein field equation, the total mass of the spacetime is at a singularity, but in most of the NaS solutions, mass is distributed around the singularity. The mass function could be written as,
\begin{equation}
    \mathcal{J}(r) = \sum_{n=0}^{\infty} a_{n} r^{n\epsilon} =a_{0} + a_{1} r^{\epsilon} + a_{2} r^{2\epsilon} +......,
\end{equation}
where $\epsilon = \pm 1$, for finite boundary spacetime $\epsilon = 1$, for asymptotically flat spacetime, $\epsilon = -1$.
In the case of a BH $a_{0} \neq 0$, however, in a NaS case $a_{0} = 0$.
In the vacuum solution of the Einstein field equation, the total mass of the spacetime is at singularity but in the NaS solutions, mass is distributed around the singularity. Hence in Schwarzschild spacetime $a_{0} = 2M$ and $\sum_{n=1}^{\infty} a_{n} r^{n\epsilon} = 0$. At the center of a massive celestial body, there are several possibilities, like a BH, NaS, gravastar, and other compact objects. Two possibilities are discussed here: BHs and naked singularities.

\subsection{Matching condition for finite boundary spacetime}
A smooth connection between two spacetimes at a timelike or spacelike hypersurface requires satisfying two junction conditions on the corresponding hypersurface \cite{refn1}. First, both sides of the matching hypersurface ($\Sigma$) must have the same induced metric. Second, in both the internal and external spacetimes, the extrinsic curvature of the matching hypersurface needs to match. The extrinsic curvature is expressed as: 
\begin{equation}
    K_{ab} = e^{\alpha}_{a}e^{\beta}_{b}\nabla_{\alpha}\eta_{\beta},
\end{equation}
where $\eta^{\beta}$ and $e^{\alpha}_{a}$ are the normal and tangents to the hypersurfaces, respectively. 

For the generic cases of gluing two Lorentzian manifolds, $\mathcal{M}_{int}$ (collapsing body) and $\mathcal{M}_{ext}$ (exterior metric), with a smooth continuous manifold, both conditions are necessary and sufficient. The process involves joining spacetimes via thin shell formalism across a boundary hypersurface denoted as $\Sigma = \partial \mathcal{M}_{int} \cap \partial\mathcal{M}_{ext}$. The total Lorentzian manifold could be $\mathcal{M} = \mathcal{M}_{int} \cup \mathcal{M}_{ext}$ \cite{Israel:1966rt}. The outside matching metric is considered to be a Schwarzschild metric, which is a vacuum solution of the Einstein field equations. Thus, we can write the external Schwarzschild metric as,
\begin{equation}
    ds^2 = -\left(1-\frac{2M_{ADM}}{R_{b}}\right)dt^2 + R_{b}^2 d\Omega^2 .\label{schmatch}
\end{equation}
Here, $M_{ADM} = GM/c^2$ is the ADM mass of spacetime, and $R_{b} $ is the boundary radius of spacetime. Matching Eq.~\eqref{schmatch} with Eq.~\eqref{genmetric} at the hypersurface $r = R_{b}$, for the zero radiation case, we obtain:
\begin{equation}
    r_{int} = r_{ext} = R_{b},
\end{equation}
\begin{equation}
    \mathcal{J}(R_{b}) = 2M_{ADM}, \label{adm}
\end{equation}
\begin{equation}
    \ddot{r} = -\frac{\mathcal{J}(R_{b})}{2r_{int}^2} = -\frac{M_{ADM}}{r_{ext}^2}.
\end{equation}
Here, $r_{int}$ and $r_{ext}$ are coordinate radii of internal and external spacetime, respectively. 

\begin{figure*}
    \centering
\subfigure[]
{\includegraphics[width=6.5cm]{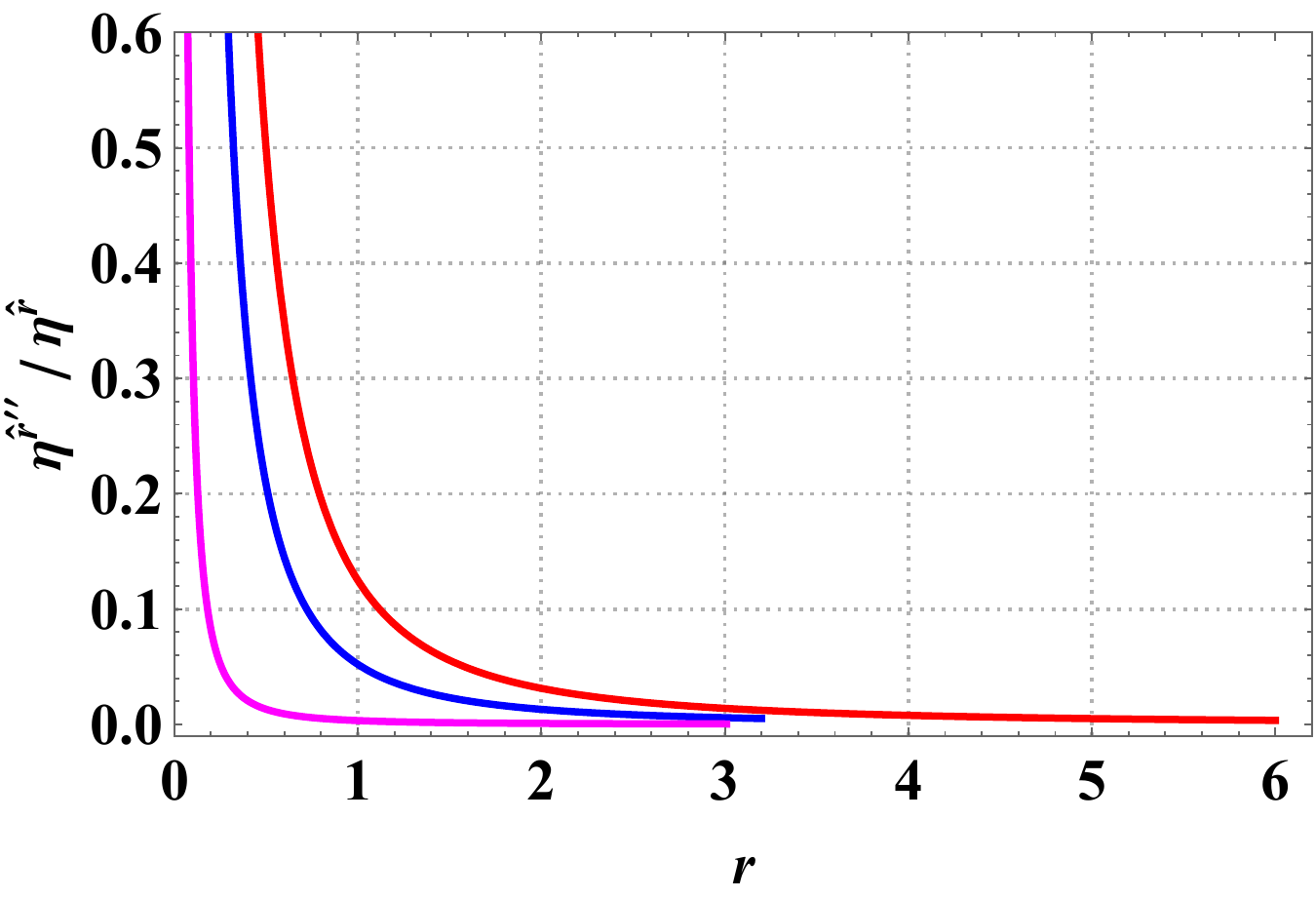}\label{fig:t=0}}
\hspace{0.1cm}
\subfigure[]
{\includegraphics[width=6.5cm]{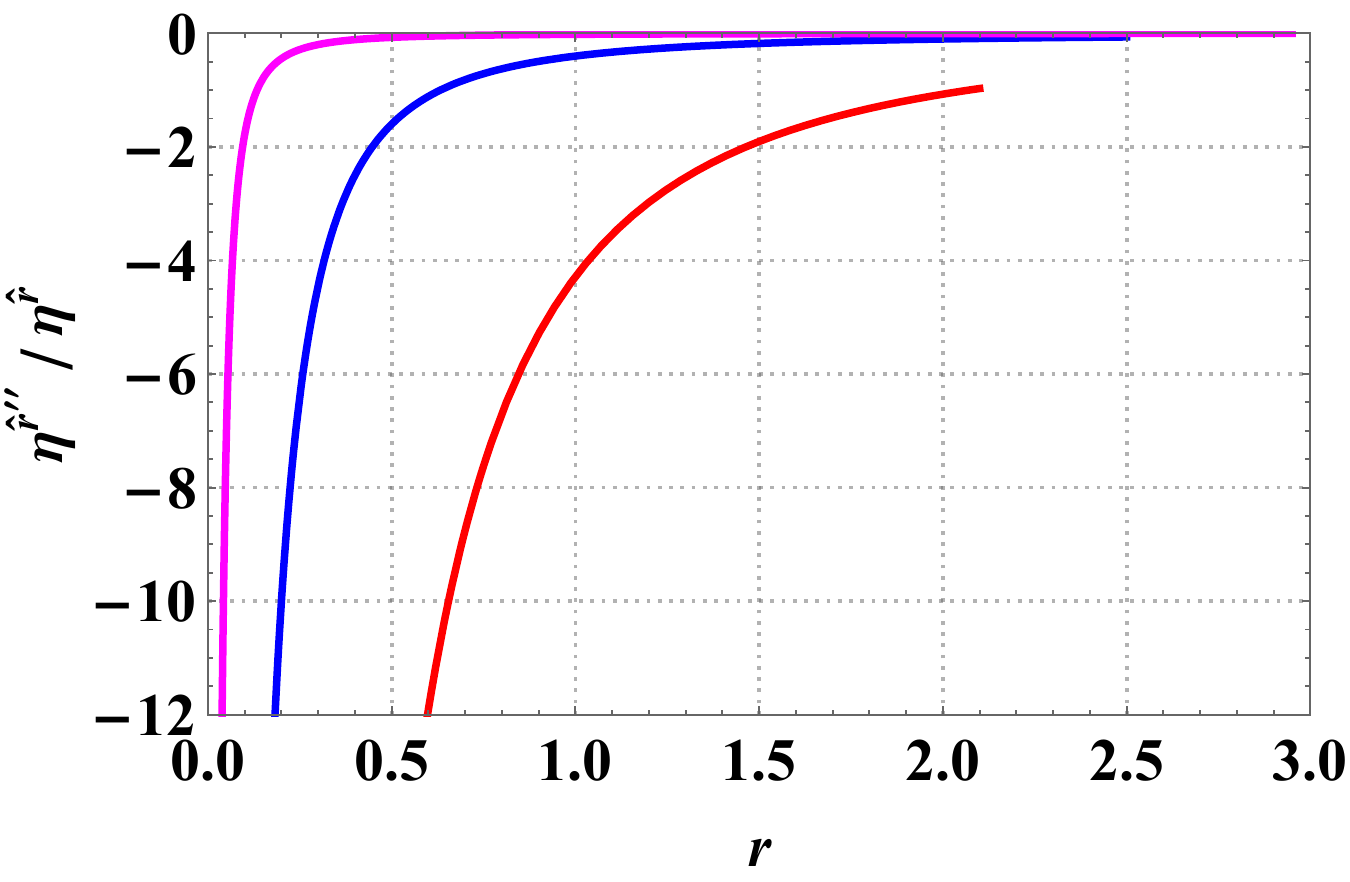}\label{fig:t=1}}\\
\subfigure[]
{\includegraphics[width=6.5cm]{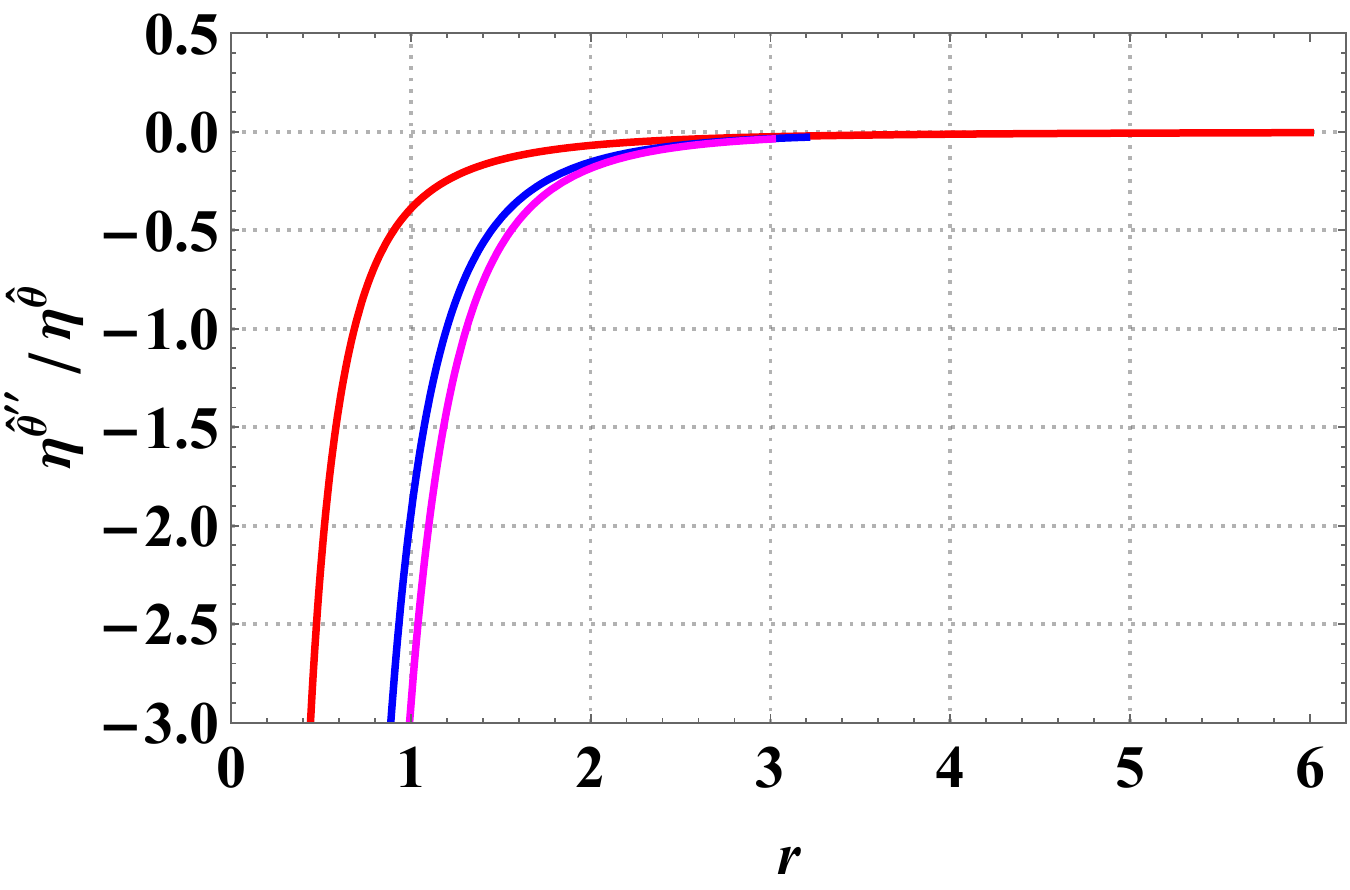}\label{fig:t=3}}
\hspace{0.1cm}
\subfigure[]
{\includegraphics[width=6.5cm]{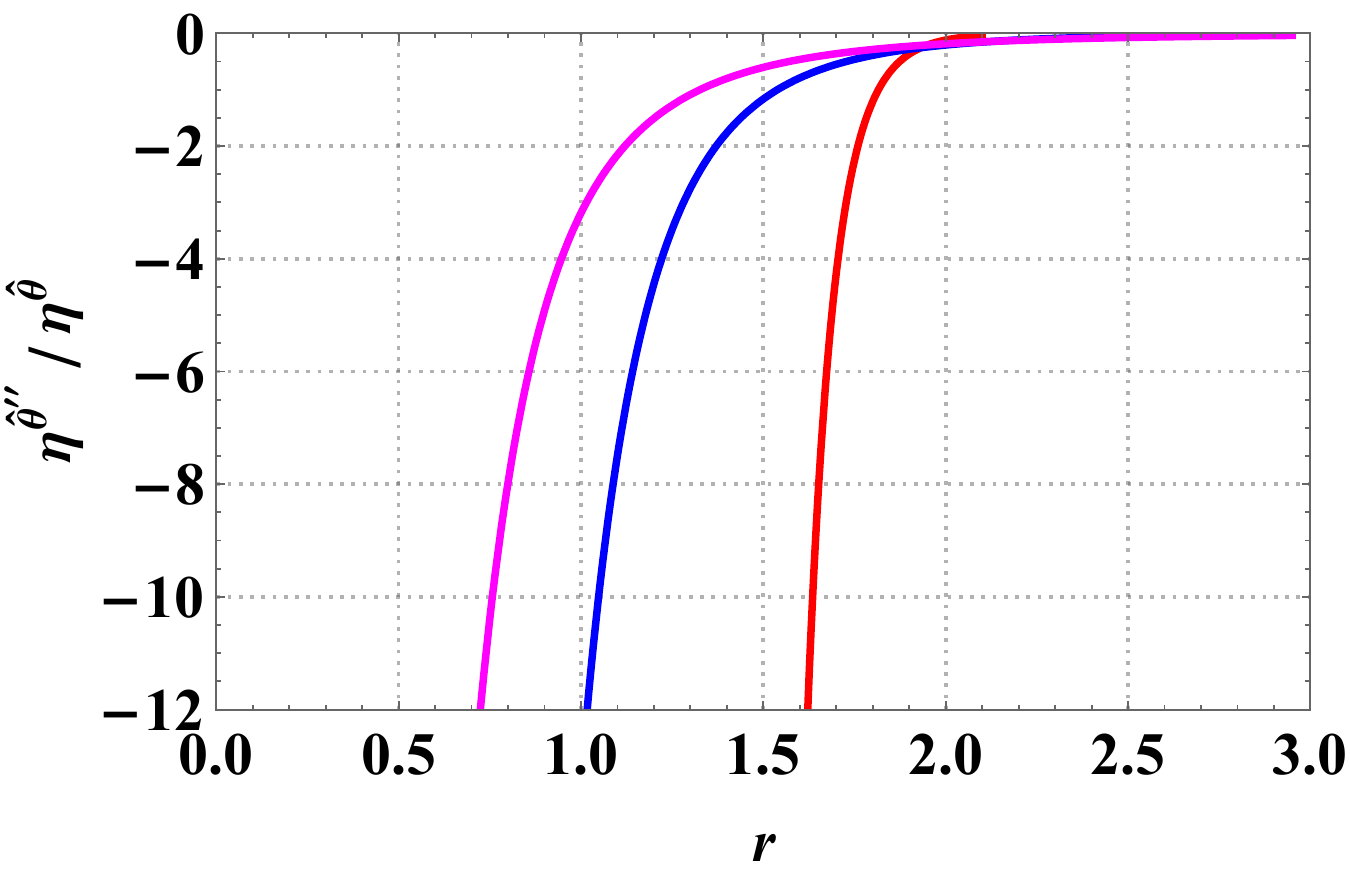}\label{fig:t=4}}\\
\subfigure[]
{\includegraphics[width=6.5cm]{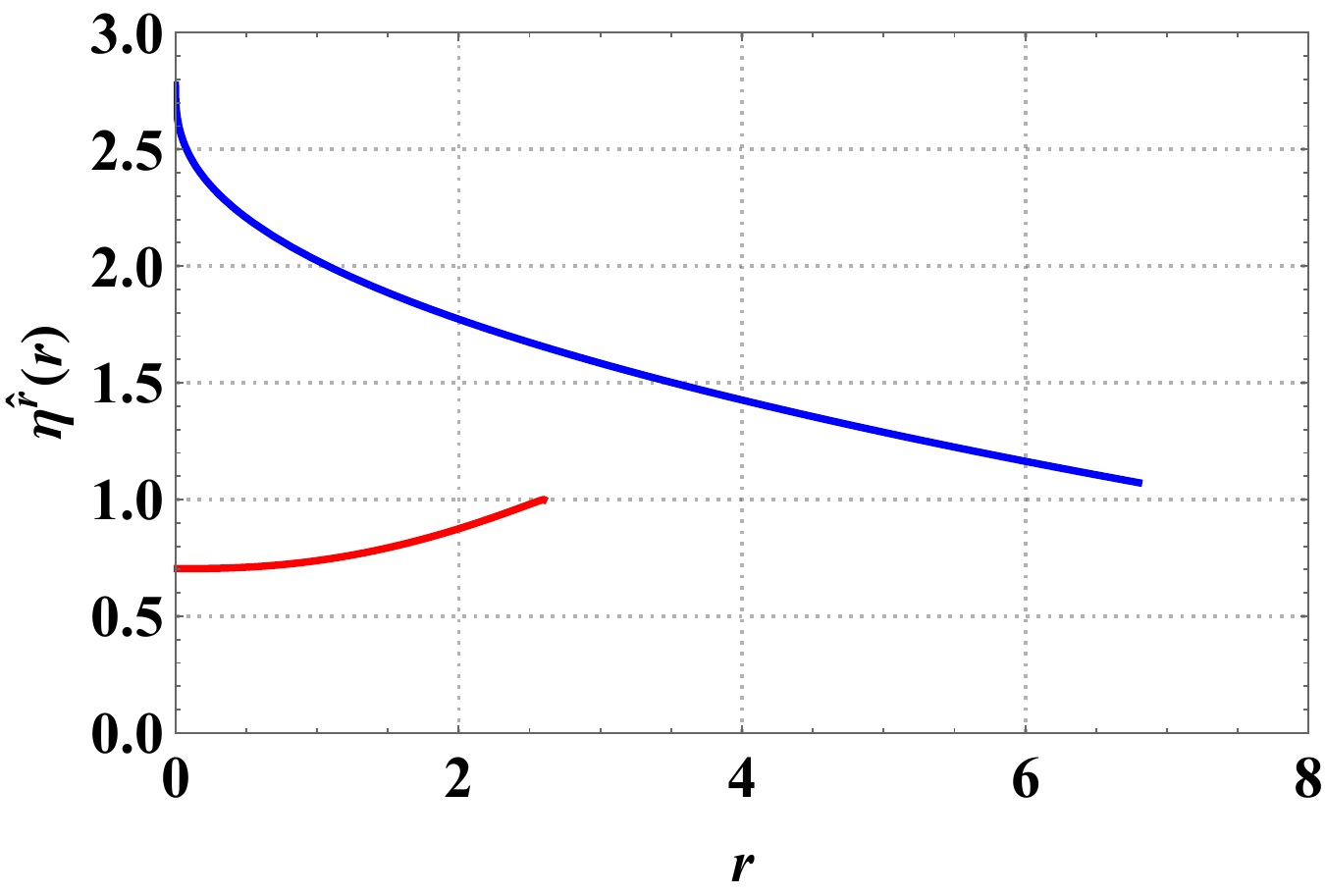}\label{jacobiradial}}
\hspace{0.1cm}
\subfigure[]
{\includegraphics[width=6.0cm]{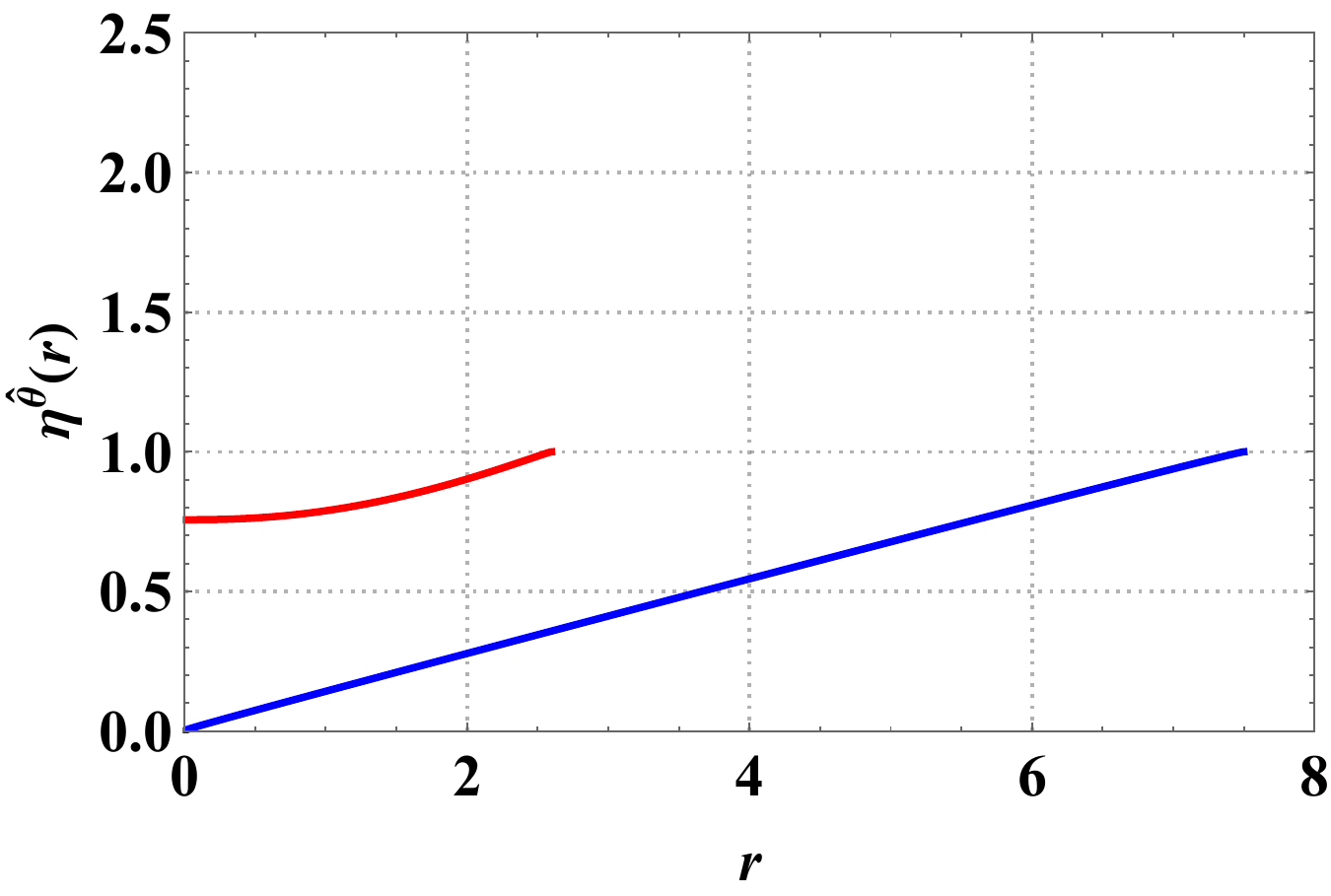}\label{jacobiangular}}
 \caption{This figure shows the radial and angular tidal force components ($\eta^{\hat{r}''} / \eta^{\hat{r}}$) and ($\eta^{\hat{\theta}''} / \eta^{\hat{\theta}}$) versus coordinate radius $(r)$, respectively. Figures (a) and (b)show the radial tidal force for $M_{0}<2/3$ and $M_{0}>2/3$, respectively.
 Similarly, figures (c) and (d)show the angular tidal force for $M_{0}<2/3$ and $M_{0}>2/3$, respectively. For (a) red, blue, and magenta lines show $R_{b} = 6M$, $R_{b} = 3.2M$ and, $R_{b} = 3.01M$, respectively. For (b)  red, blue, and magenta lines show $R_{b} = 2.1M$, $R_{b} = 2.5M$, and $R_{b} = 2.95M$, respectively. (c) for red $R_{b} = 6M$ and $r_{0} = 5.5M$, for blue $R_{b} = 3.2M$ and $r_{0} = 3.0M$, for magenta line $R_{b} = 3.01M$ and $r_{0} = 3.0M$. (d) for red $R_{b} = 2.1M$ and $r_{0} = 2.0M$, for blue $R_{b} = 2.5M$ and $r_{0} = 2.4M$, for magenta line $R_{b} = 2.95M$ and $r_{0} = 2.9M$. In the above figures, the radial ($\eta^{\hat{r}}$) and angular ($\eta^{\hat{\theta}}$) components of the Jacobi field versus coordinate radius for JMN1 spacetime are shown. Fig.~(\ref{jacobiradial}) illustrates how the radial components of the Jacobi field change with coordinate radius $r$. Fig.~(\ref{jacobiangular}) demonstrates how the angular components of the Jacobi field change with coordinate radius $r$. The initial conditions chosen are $\eta^{\hat{r}} = 1$, $\eta^{\hat{\theta}} = 1$, $\dot{\eta}^{\hat{r}} = 0$, and $\dot{\eta}^{\hat{\theta}} = 0$, where the overdot represents the derivative with respect to proper time $\tau$. Red and blue lines show for $R_{b} =  r_{0} = 7.5M$ and $R_{b} = r_{0} = 2.6M$, respectively. Here we take $M = 1$.}
 \label{fig:1}
\end{figure*}

\subsection{Joshi-Malafarina-Narayan type 1 Naked Singularity}
In Eq.~(\ref{g}), if we take the mass function $\mathcal{J}(r) = M_{0}r$ and radial pressure $p_{r} = 0$ with a finite integral over $r$ to $R_{b}$, and in Eq.~\ref{g} for $f(r)$ function, with matching condition $\mathcal{J}(R_{b}) = 2GM/c^2 = M_{0} R_{b}$. We can derive the line element of the JMN1 spacetime,
\begin{eqnarray}
 ds^2=-(1-M_0) \left(\frac{r}{R_b}\right)^\frac{M_0}{1-M_0}dt^2 + \frac{dr^2}{1-M_0} + r^2d\Omega^2\,\, \nonumber,
\end{eqnarray}
where the dimensionless parameter $M_0$ should be $0<M_0<1$ and $R_b$ is boundary radius of JMN1 spacetime that matches with the external Schwarzschild spacetime. It is demonstrated in \cite{psJoshi1} that JMN1 spacetime can form as an endstate of the gravitational collapse of an anisotropic matter fluid in asymptotic time. At the coordinate center, this spacetime has a null singularity for $M_{0}>2/3$ and a timelike singularity when $M_{0}<2/3$ \cite{Bambhaniya:2021jum}. JMN1 NaS solution is an anisotropic fluid solution of Einstein equations, where the energy density and pressures can be expressed as,
\begin{eqnarray}
\rho = \frac{M_0}{r^2},\,\, P_r = 0,\,\, P_\theta = \frac{M_0^2}{4(1-M_0)r^2}\,\, .
\end{eqnarray} 
In \cite{Goel:2015zva}, tidal force in JMN1 spacetime is examined. However, they do not show the Jacobi field or the tidal disruption radius in JMN1 spacetime. JMN1 spacetime is astrophysically relevant. According to the EHT publication \cite{EventHorizonTelescope:2022xqj}, JMN1 spacetime is a potential candidate for the galactic center. There is a significant amount of literature discussing particle motions \cite{Bambhaniya:2019pbr} in JMN1 spacetime and the shadow cast by them \cite{Joshi2020}, and it is demonstrated that these findings are critical to understanding the distinguishing observable signatures of BHs and naked singularities.  

\section{Jacobi fields on timelike radial geodesics in static spherically symmetric (SSS) spacetimes}\label{appB}
Our focus lies in locally maximum curves, and we take into account radial SSS spacetimes. The geodesic deviation equation for the timelike radial geodesics is derived in this section; the spacetime under consideration determines the precise form of the equation and, consequently, the Jacobi vector field. First, we derive the equations for every timelike geodesic. Using Eq.~(\ref{rdot}), we can write Eq.(\ref{tidalg}) and Eq.~(\ref{tidalg1}) as, 

\begin{eqnarray}\label{n36}
 \frac{(e^2 - f(r))}{f(r)g(r)}\frac{d^2 \eta^{\hat{r}}}{dr^2} & + & \frac{g'(r)f(r)^2 - e^2 (f'(r)g(r)+g'(r)f(r))}{2g(r)^2 f(r)^2}\frac{d\eta^{\hat{r}}}{dr} -\nonumber\\
 & & \frac{ f(r) f'(r) g'(r) + g(r)(f'(r)^2 - 2f(r)f''(r))}{4 f(r)^2  g(r)^2}\eta^{\hat{r}} =0,
 \end{eqnarray}
 \begin{eqnarray}\label{n37}
 \frac{(e^2 - f(r))}{f(r)g(r)}\frac{d^2 \eta^{\hat{i}}}{dr^2} & + & \frac{g'(r)f(r)^2 - e^2 (f'(r)g(r)+g'(r)f(r))}{2g(r)^2 f(r)^2}\frac{d\eta^{\hat{i}}}{dr} -\nonumber\\
 & & \frac{g'(r)f(r)^2 - e^2 \left(g(r)f'(r)+ f(r)g'(r)\right)}{2r g(r)^2 f(r)^2} \eta^{\hat{i}} =0,
\end{eqnarray}

\subsection{Tidal force and Jacobi field in the JMN1 Naked Singularity spacetime}
We consider a simple example, $\mathcal{J}(r) = M_{0}r$ which will give a solution of JMN1 spacetime given in \cite{psJoshi1}. Radial tidal force in JMN1 spacetime is given as follows,
\begin{equation}
    \frac{d^2 \eta^{\hat{r}}}{d\tau^2}= \frac{M_{0}(2 - 3 M_{0})}{4(1 - M_{0}) r^2}\eta^{\hat{r}}, \label{tidaljmn}
\end{equation}
Here, $M_{0}>2/3$ shows the compressive nature of radial tidal force. While $M_{0}<2/3$ shows the stretching nature of radial tidal force. In \cite{Bambhaniya:2021jum}, we show by using the Penrose diagram that for $M_{0}>2/3$, the gravitational center has a null singularity. While in $M_{0}<2/3$ have timelike singularity. Therefore, close to the null singularity radial tidal is infinitely compressive, and in timelike singularity, it shows infinitely stretching.

Angular tidal force in JMN1 spacetime is given as follows,
\begin{equation}
\frac{d^2 \eta^{\hat{i}}}{d\tau^2}= -\frac{E M_{0}}{2(1-M_{0})r^2}\left(\frac{R_{b}}{r}\right)^{\frac{M_{0}}{1 -M_{0}}} \eta^{\hat{i}}, \label{tidaljmn1}
\end{equation}
where dimensionless energy $e$ can be calculated using the initial conditions, at $r = r_{0}$ radial velocity $\dot{r} = 0$. Therefore,
\begin{equation}
     e^2 = (1 - M_{0})\left(\frac{r_{0}}{R_{b}}\right)^{\frac{M_{0}}{1-M_{0}}}
\end{equation}
Hence Eq.~(\ref{tidaljmn1}) can be written as,
\begin{equation}
    \frac{d^2 \eta^{\hat{i}}}{d\tau^2}= -\frac{ M_{0}}{2r^2}\left(\frac{r_{0}}{r}\right)^{\frac{M_{0}}{1 -M_{0}}} \eta^{\hat{i}}, \label{tidaljmn12}
\end{equation}
Here for both cases, null singularity and timelike at the gravitational center show compressive forces.

\end{document}